\documentstyle[prd,aps,floats]{revtex}

\begin{document}
\draft

\input epsf 

\renewcommand{\topfraction}{0.8}
\twocolumn[\hsize\textwidth\columnwidth\hsize\csname
@twocolumnfalse\endcsname

\title{Resonance enhancement of particle production during reheating}
\author{G. Palma and V. H. C\'{a}rdenas}
\address{Departamento de f\'{\i}sica, Universidad de Santiago de Chile, Casilla 307,\\
Correo 2, Santiago}
\date{\today}
\maketitle

\begin{abstract}
We found a consistent equation of reheating after inflation which shows that
for small quantum fluctuations the frequencies of resonance are slighted 
different from the standard ones. Quantum interference is taken into account and we found that at large fluctuations the process mimics 
very well the usual parametric resonance but proceed in a different
dynamical way. The analysis is made in a toy quantum mechanical model and 
we discuss further its extension to quantum field theory.
\end{abstract}

\pacs{PACS numbers: 98.80.Cq \hfill USACH preprint 00/13 }

\preprint{USACH-00/13}

\vskip2pc]


Inflation has become one of the most successful ideas of high energy
cosmology today \cite{Linde}. Actually most of the current work on
experimental cosmology are based on ideas, or designed to test inflationary
models \cite{rmp}. No matter which inflationary model we assume, at the end
of this period, it must reheat the universe. This period,
called reheating, is very important because basically all elementary
particles populating the universe were created. Much of the current interest
in reheating follows after the discovery of an exponential amplification in
the number of the particles produced during preheating. Source of this
exponential growth is the presence of \ the parametric and stochastic
resonance \cite{TB,KLS,KLS1,STB,BHdV} in the solutions of the equations of
motion. Recent investigations address the problem of parametric
amplification of super-horizon perturbations \cite{suphub} during preheating
and their consequences in the CMB spectrum \cite{pertur}.

Because the study of models of interacting fields evolving along the
expansion of the universe is a highly non-trivial issue, analytical progress
in this direction has been difficult to obtain. However, as we describe
later, most of the main features of the observed phenomena can be obtained
by studying a toy quantum mechanical model of two bi-quadratically coupled
oscillators. Furthermore the recent interest in amplification of
fluctuations can be described in this context.

In this letter, we describe in a simple way how parametric and stochastic
resonance arise from the dynamics of a system of two coupled oscillators and
we also identify the main problems that occur with the standard approaches,
such as the energy conservation problem. We derive further in a consistent
way, a semiclassical equation of motion for the scalar condensate field $%
R(t) $ (see Eq.(\ref{leR}) below). From the analysis of this equation, it is
shown that the preheating phase proceeds in a different dynamical way,
consistent with the Heisenberg's principle. This fact also leads to
appropriate initial conditions - an initially non-zero value for the
fluctuations - to allow an efficient energy transfer between the oscillators.
The equations obtained (Eq.(\ref{leX}) and Eq.(\ref{leR})) can be considered
as a classical system, allowing us to perform a complete numerical study 
\cite{JC}. A first numerical study shows a novel saturation effect in
the small $R$-region of the equation for the fluctuation. This effect holds
at the beginning of the energy transfer between the oscillators. We extend
our results to quantum field theory, and take into account the
expansion of the universe. This is straightforward, and the results
obtained in our previous analysis can be extended with slight modifications.

The model we consider is a classical oscillator $X$ interacting with a
purely quantum mechanical oscillator $\hat{Y}$ described by the Lagrangian

\begin{equation}
L=\frac{1}{2}\dot{X}^{2}+\frac{1}{2}\dot{\hat{Y}}^{2}-\frac{1}{2}\omega ^{2}X^{2}-%
\frac{1}{2}\Omega ^{2}\hat{Y}^{2}-\frac{1}{2}g^{2}X^{2}\hat{Y}^{2},  \label{lagra}
\end{equation}
where $X(t)$ corresponds to the classical inflaton field $\phi (t)$ and $%
\hat{Y}(t) $ the created scalar field $\hat{\chi}(x)$ of inflationary cosmology
respectively. For reasons of simplicity we will first neglect the expansion
of the universe, and discuss at the end the effects and possible
modifications of including such a friction term. The equations of motion are
then corresponding to two coupled harmonic oscillators with time-dependent
frequencies

\begin{equation}
\ddot{X}+(\omega ^{2}+g^{2}\hat{Y}^{2})X=0,  \label{lagx}
\end{equation}

\begin{equation}
\ddot{\hat{Y}}+(\Omega ^{2}+g^{2}X^{2})\hat{Y}=0.  \label{lagy}
\end{equation}
Let us now introduce time-independent creation and annihilation Heisenberg
operators $a^{\dagger }$ and $a$ respectively using the Ansatz

\begin{equation}
\hat{Y}=f(t)a+f^{\ast }(t)a^{\dagger },  \label{rheis}
\end{equation}
where, due to the standard commutator $\left[ a,a^{\dagger }\right] =1$ for
creation-annihilation operators, $f(t)$ satisfies the Wronskian condition

\begin{equation}
\dot{f}^{\ast }(t)f(t)-f^{\ast }(t)\dot{f}(t)=-i.  \label{comf}
\end{equation}
Inserting the Ansatz of Eq. (\ref{rheis}) into Eq. (\ref{lagy}), we find
that $f(t)$ satisfies the equation of motion

\begin{equation}
\ddot{f}(t)+(\Omega ^{2}+g^{2}X^{2}(t))f(t)=0,  \label{mateqf}
\end{equation}
where the normalization is fixed by the condition (\ref{comf}). If we
consider $\hat{Y}$ in the frame of quantum field theory, Eq. (\ref{mateqf}) would
be the equation for the zero-momentum term of its Fourier expansion.
Motivated by the plane wave expansion of a quantized-field operator we
use the following Ansatz for the function-coefficient $f(t)$

\begin{equation}
f(t)=\frac{1}{\sqrt{2W(t)}}\exp \left( -i\int^{t}W(t^{\prime })dt^{\prime
}\right) ,
\end{equation}
which satisfies automatically Eq. (\ref{comf}). Inserting this Ansatz into
Eq. (\ref{mateqf}), we obtain the nonlinear differential equation for $W(t)$

\begin{equation}
\frac{1}{2}\frac{\ddot{W}}{W}-\frac{3}{4}\left( \frac{\dot{W}}{W}\right)
^{2}+W^{2}=m^{2}(t),  \label{weqt}
\end{equation}
where $m^{2}(t)=\Omega ^{2}+g^{2}X^{2}(t)$. Eq. (\ref{weqt}) describes a
non-linear dissipative differential equation for $W(t)$. Instead of solving
this rather cumbersome equation numerically, we will consider the Hamiltonian
associated to the Lagrangian of Eq. (\ref{lagra}) and construct a classical
effective Hamiltonian by projecting onto the subspace of the number operator
$\hat{n}=a^{\dagger}a$, generated by $\{\left| 4\right\rangle ,\left| 3\right\rangle,\left| 2\right\rangle ,\left| 1\right\rangle ,\left| 0\right\rangle \}$,
which we will call $S_{5}$. The vacuum is as usual defined by the condition
$a\left| 0\right\rangle =0$.
The choice of the non-trivial subspace $S_{5}$ is two fold. First, it attempts
a fully quantum-mechanical treatment of this problem, allowing the possibility
of quantum interference and other quantum effects in the effective classical
theory, and secondly it allows an exact analitical solution of the quantum sector
of the model.
We first define the classical variable $R^{2}(t)=\left\langle 0\left| \hat{Y}^{2}\right| 0\right\rangle $.
The operator $\hat{Y^{2}}$ reads

\[
\left[ \hat{Y}^{2}\right] _{nm}=\left[ 
\begin{array}{ccccc}
\left| f\right| ^{2} & 0 & \sqrt{2}f^{2} & 0 & 0 \\ 
0 & 3\left| f\right| ^{2} & 0 & \sqrt{6}f^{2} & 0 \\ 
\sqrt{2}f^{\ast 2} & 0 & 5\left| f\right| ^{2} & 0 & 2\sqrt{3}f^{2} \\ 
0 & \sqrt{6}f^{\ast 2} & 0 & 7\left| f\right| ^{2} & 0 \\ 
0 & 0 & 2\sqrt{3}f^{\ast 2} & 0 & 9\left| f\right| ^{2}
\end{array}
\right] . 
\]
and can be diagonalized in $S_{5}$ explicitly $\left[ \hat{Y}^{2}\right] _{nm}V_{i}=\lambda _{i}V_{i}$,
where $V_{i}$ and $\lambda_{i}$ are the eigenvectors and eigenvalues
respectively. Because of the similar structure of both operators,
$\hat{Y}^2$ and $\hat{P}_{Y}^2=\dot{\hat{Y}}^{2}$, we can use the Ansatz
of Eq. (\ref{rheis}), and the effective Hamiltonian can be diagonalized
exactly as

\begin{equation}
H_{ij}^{eff}=\frac{1}{2}[P_{X}^{2}+P_{R}^{2}+\frac{1}{4R^{2}}+\omega
^{2}X^{2}+m^{2}(t)R^{2}]y_{i}\delta _{ij},  \label{L_eff}
\end{equation}
where $i,j=0,..4$. Up to irrelevant constants, this Hamiltonian describes the effective dynamics of two classical variables $X$ and $R$, which corresponds to the original classical and quantum oscillators. From Eq. (\ref{L_eff}) the equations of motion are

\begin{equation}
\ddot{X}+(\omega ^{2}+g^{2}R^{2})X=0,  \label{leX}
\end{equation}

\begin{equation}
\ddot{R}+(\Omega ^{2}+g^{2}X^{2}(t))R-\frac{1}{4R^{3}}=0.  \label{leR}
\end{equation}
The above equations actually correspond to two classical coupled
particles-like degrees of freedom. Note that the centrifugal term keeps the
fluctuation $R$\ away from zero, consistent with Heisenberg's principle and
giving us appropriated natural initial conditions for numerical studies \cite
{JC}. These equations are very simple and represent the starting point of
our analysis.

A similar derivation of the classical Hamilton dynamics of a classical
oscillator interacting with a quantum mechanical oscillator has been made in 
\cite{Cooper}, in the context of semiquantum chaos. Nevertheless, the chosen
classical-variable parametrization leads to cumbersome nonlinear dissipative
differential equations.

We note here, that backreaction is taken into account in Eq. (\ref{leX}),
through the dependence on the classical condensate variable $R$. We start
the discussion of the classical dynamics by considering the perturbative
regime of the equations of motion, at small coupling constant $g^{2}\ll 1$.
For $g^{2}=0$, a solution of Eq. (\ref{leX}) is $X(t)=X_{0}\sin (\omega t)$.
Inserting this expression into Eq. (\ref{mateqf}) we obtain a Mathieu
equation

\begin{equation}
f^{\prime \prime }(z)+\left[ A(q)-2q\cos 2z\right] f(z)=0,  \label{emath}
\end{equation}
where $z=\omega t$, $q=g^{2}X_{0}^{2}/4\omega ^{2}$ and $A(q)=\Omega
^{2}/\omega ^{2}+2q$. This equation is well-known and has been used to
discuss different mechanisms of particle production during the reheating
period of the universe, such as parametric and stochastic resonance \cite{TB}%
-\cite{STB}. In fact, for $2q<A$ this equation leads to narrow parametric resonance 
\cite{Landau} and for $2q>A$ leads to broad parametric resonance \cite{KLS}
(called stochastic resonance if expansion of the universe is considered). From
the definitions above for $A$ and $q$, we observe that only the first region
is allowed. The limit curve $A=2q$ in the space of parameters can be
discussed by considering the limit case $\Omega ^{2}=0$ of the equivalent
equation

\begin{equation}
\ddot{f(t)}+(\Omega ^{2}+g^{2}X_{0}^{2}\sin ^{2}(\omega t))f(t)=0. \label{trece}
\end{equation}
This equation leads also to a linear growth of the amplitude $f(t)$, close
to the minimum of the potential for $X(t)$. This can be explained
intuitively because this equation reads $\ddot{f}(t)=0$ at these points (see
Fig. 1), and hence has solutions which raise linearly with time.

\begin{figure}[t]
\centering \leavevmode \epsfysize=6.0cm \epsfbox{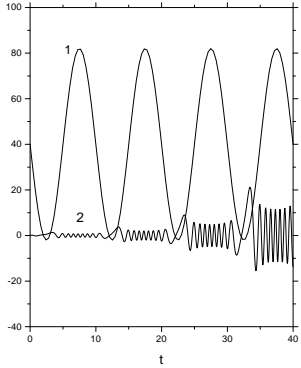}
\caption[fig1]{\label{fig:osc1} We show the oscillations of the source frequency 
(see Eq. (\ref{trece})) which is responsible for parametric resonance (curve 1) 
and the response $f(t)$ (curve 2). Note that when the $X$ oscillations pass through 
zero, the maximun amplifications occur. }
\end{figure}

\begin{figure}[t]
\centering \leavevmode \epsfysize=6.0cm \epsfbox{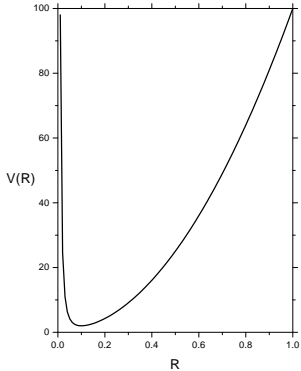}
\caption[fig1]{\label{fig:osc2} We plot the potential of Eq. (\ref{potR}) in a convenient scale. Note the the potential has a wall close to origen $R=0$.}
\end{figure}

\begin{figure}[t]
\centering \leavevmode \epsfysize=6.0cm \epsfbox{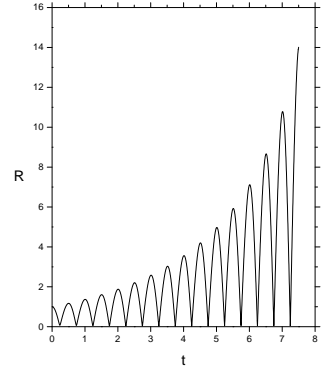}
\caption[fig1]{\label{fig:osc3} This plot shows the parametric amplification of $R$. 
In the context of a quantum field theory, this result is interpreted as a particle 
production process.}
\end{figure}

\begin{figure}[t]
\centering \leavevmode \epsfysize=6.0cm \epsfbox{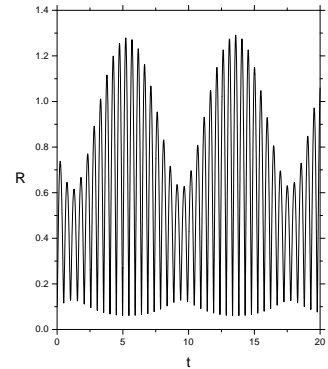}
\caption[fig1]{\label{fig:osc4} In the small amplitudes region, close to the resonance 
conditions, there is a novel saturation effect due to detuning coming from nonlinearities 
of the equation of motion. See explanation below Eq. (\ref{potexp}).}
\end{figure}

\begin{figure}[t]
\centering \leavevmode \epsfysize=6.0cm \epsfbox{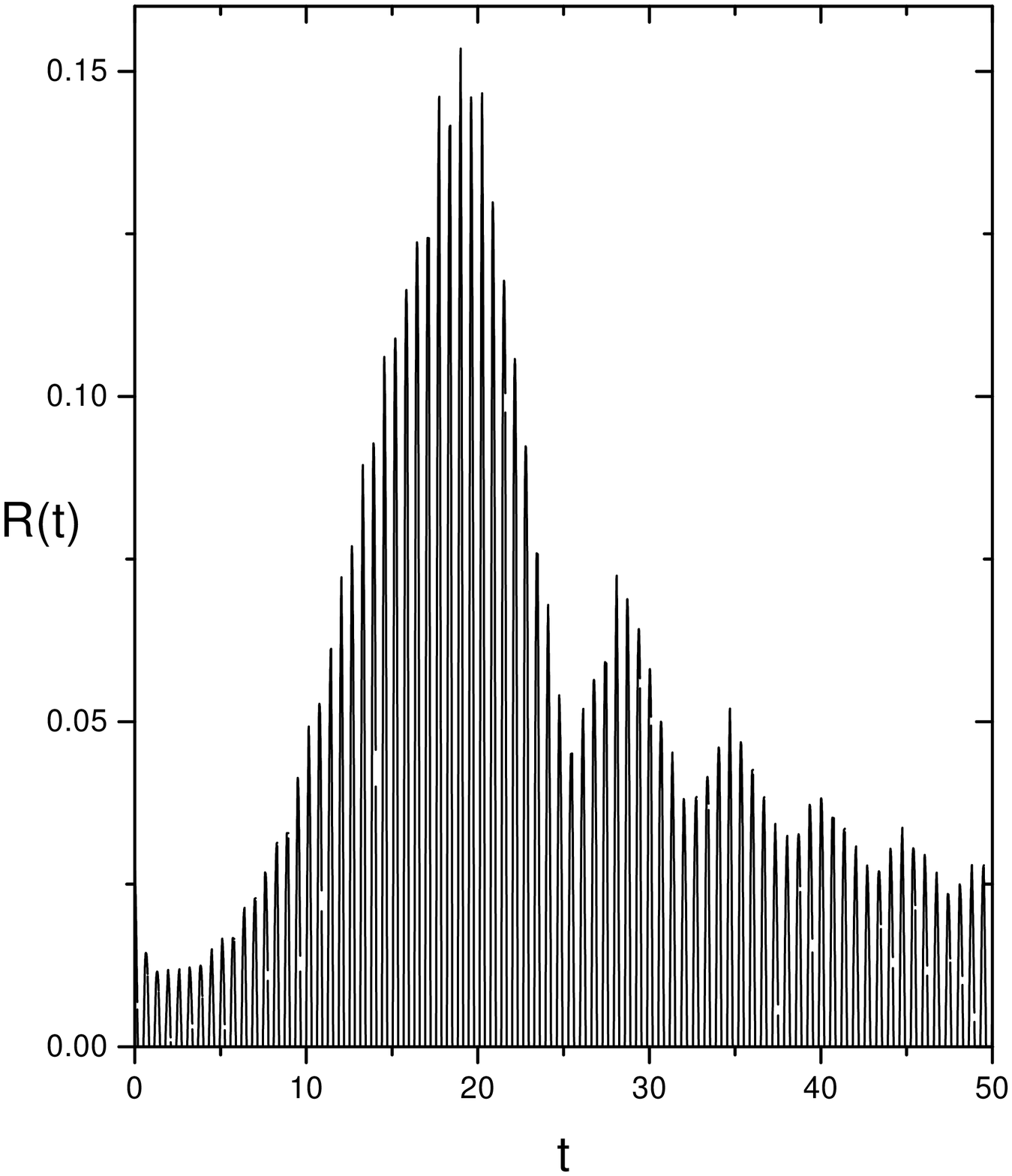}
\caption[fig1]{\label{fig:osc5} This plot shows how $R$ ceases to grow exponentially after certain time, when the 
expansion of the universe is considered. The resonance structure of the secondary
maxima is given by the detuning effect due to the presence of the $a^{2}$-factor in
the effective frequency term of Eq. (\ref{eqr}).
}
\end{figure}

However, it is natural to ask whether or not a non-exactly oscillatory form
of $X$ will lead to an exponential growth of $f(t)$. Moreover, we know from
this simple system that as soon as $f(t)$ start to growth, the functional
form of $X$ will change, not only in amplitude but also in frequency. Also,
attention must be down to solutions of Eq. (\ref{leR}) and not of Eq. (\ref
{mateqf}), because in general its solutions does not satisfy the Wronskian
condition Eq. (\ref{comf}). The best way to take into account the energy
conservation, is by studying a numerical solution \cite{JC}. Even, the
system (\ref{leX},\ref{leR}) can be seen as a classical system allowing us
to perform in a easier way such a study.

So, let us return to equations (\ref{leX}) and (\ref{leR}). The second
equation can be interpreted as a particle moving in the presence of the
potential

\begin{equation}
V(R)=\frac{1}{2}m^{2}(t)R^{2}+\frac{1}{8R^{2}}.  \label{potR}
\end{equation}
Figure 2 shows a plot of the potential for a convenient value of $m^{2}$. We
see that $R=0$ acts as an infinite wall, which divides the coordinate space
into two disconnected regions. Depending on which sign of the root we
choose, the particle will remain in one of these two regions.

In order to analyze this potential, there are two scales to be considered.
For large amplitudes $R\gg 1/\sqrt{2m}$, the potential looks essentially
like a parabola with an infinite wall at $R=0$ (see Fig. 2). The solution of
the equation without the non-linear term, which leads to the infinite wall,
has resonances at $\Omega _{n}=2\omega /n$, as in the conventional
parametric resonance (see \cite{Landau}). One could naively expect to find
resonances of the complete equation at twice this value because of the
presence of the wall at $R=0$. But this is not true. The reason for this
behavior is as follows. The solution of the problem without the wall-term is
invariant under the shift $\omega t\rightarrow \omega t+\pi $. But this
corresponds to the same solution of the system in the presence of the wall,
when the phase is shifted by $\pi $ as $R$ approaches zero. Figure 3 shows
the growth of the amplitude of oscillations for the main resonance band $%
\Omega =2\omega $. This curve has a similar shape and properties of the one
corresponding to narrow parametric resonance, Fig. 2 in \cite{KLS1}.

For small amplitudes $R\sim 1/\sqrt{2m}$, we can expand the potential around
the time dependent minimum $R_{m}(t)=\pm 1/\sqrt{2m(t)}$

\begin{equation}
V(R)\simeq \frac{1}{2}m^{2}(t)+2(R-R_{m})^{2}m^{2}(t).  \label{potexp}
\end{equation}
From (\ref{potexp}), we see that for amplitude values close to the minimum,
the potential develops a linear term, which leads to non-linear
oscillations. Such oscillations hold in an almost energy conserved
environment, because the amplitude of $R$ is small, so this effect is a
genuine one and holds during the beginning of the transfer of energy. Fig. 4
shows the typical behavior of these oscillations. There are two regions
separated by the size of the amplitude. When the amplitude is small and a
resonance condition is fulfill, the linear term drives the system to larger
values of the amplitude, which is characteristic of the broad parametric
resonance (see Fig. 3 in \cite{KLS1}). Nevertheless, when the amplitude grows
the non-linear term of the equation becomes important and breaks the
resonance tuning and the growth of the amplitude saturates, while the
amplitude starts to decrease. Most of the results obtained above can be still
applied when we consider the expansion of the universe.


In quantum field theory the process follows similar lines. If we
consider the inflaton field $\phi $ and the created field $\chi $ coupled
minimally to gravity, the Lagrangian can be written as a sum ${\cal L}={\cal %
L}_{0}(\phi )+{\cal L}_{0}(\chi )+{\cal L}_{I}(\phi ,\chi ),$ where 
\begin{equation}
{\cal L}_{0}(\phi )=\sqrt{-g}[\frac{1}{2}(\partial \phi )^{2}-\frac{1}{2}%
m_{\phi }^{2}\phi ^{2}],  \label{lagqft}
\end{equation}
where the interaction is ${\cal L}_{I}=-\frac{1}{2}g^{2}\phi ^{2}\chi ^{2}$.
Let us consider $\phi (t)$ and $\chi (t)$ as homogeneous fields, but the
former as a {\it classical} and the later as a {\it quantum} operator field.
Assuming a flat FRW metric, the equations of motion obtained from the
Lagrangian (\ref{lagqft}) are 
\begin{equation}
\chi ^{\prime \prime }+2{\cal H}\chi ^{\prime }+a^{2}(M_{\chi
}^{2}+g^{2}\phi ^{2})\chi =0,  \label{eomxi}
\end{equation}
\begin{equation}
\phi ^{\prime \prime }+2{\cal H}\phi ^{\prime }+a^{2}(m_{\phi
}^{2}+g^{2}\chi ^{2})\phi =0,  \label{eomfi}
\end{equation}
where $a$ is the scale factor, a prime $^{\prime }$ means a derivative
respect to conformal time $\eta =\int dt/a(t)$, and ${\cal H}=a^{\prime }/a$%
. Defining the conformal fields as $X=a\phi $ and $\hat{Y}=a\chi $, we find the
equations 
\begin{equation}
X^{\prime \prime }+(a^{2}m_{\phi }^{2}+g^{2}\hat{Y}^{2}-\frac{a^{\prime \prime }}{a%
})X=0,  \label{xce}
\end{equation}
\begin{equation}
\hat{Y}^{\prime \prime }+(a^{2}M_{\chi }^{2}+g^{2}X^{2}-\frac{a^{\prime \prime }}{a%
})\hat{Y}=0.  \label{yce}
\end{equation}
If we neglect the coupling between $\chi$ with $\phi$ and during the 
oscillations of the inflaton, the universe expands as a one dominated  
by matter. Under such circunstances the last term inside the brackets 
can be dropped out. Moreover, we have checked numerically that the 
last term $a^{\prime \prime }/a$ can be neglected, even if the coupling 
is considered.

The semiclassical approximation leads to the equation 
\begin{equation}
R^{\prime \prime }+(a^{2}M_{\chi }^{2}+g^{2}X^{2}-\frac{a^{\prime \prime }}{a%
})R-\frac{1}{4R^{3}}=0,  \label{eqr}
\end{equation}
and similarly the corresponding for $X$ (Eq.(\ref{xce})) but replacing $\hat{Y}^{2}$ by $R^{2}$.
If we consider the corrections beyond the long-wavelenght limit of the field
$\hat{Y}$, an infinite set of non-linear coupled differential equations for
the modes $R_{k}$ prevents an analogous explicit description as the one given
by Eq. (\ref{eqr}). Nevertheless, this observation allows a perturbative treatment
of this problem by a gradient expansion of the fields, but this is beyond
the scope of our analysis.
Because the fields are scaled by $a$, the expansion of the universe should damp
the oscillations of the created field $R$ (or $\chi $) in Fig. 1 and Fig. 3.
In Fig. 5 the combined effects of backreaction and the expansion of the universe 
in the evolution of the field 
$R$ is shown. As a consequence, $R$ ceases to grow exponentially, after certain 
time. The resonance structure of the secondary maxima can be explained by the 
detuning effect due to the presence of the scale factor in the effective frequency 
term of Eq. (\ref{eqr}).


In this letter we have described a model to study reheating after inflation.
We have showed that a simple model of two coupled oscillators can reproduce
the main resonances effects, responsible for particle creation during
preheating. In addition we have derived a semiclassical equation describing
particle production consistent with Heisenberg's uncertainty principle. From
this equation we have also found that at small amplitude of the
oscillations, i. e. when the energy transfer starts, there is no resonance
amplification. This seems to indicate that there is an energy threshold to
get efficient particle production.


\section*{Acknowledgments}

G. P. would like to thank to R. Labb\'{e} for helpful discussions. G. P. was
supported in part by the projects FONDECYT 1980608 and DICYT 049631PA. V. H.
C. would like to thank CONICYT for financial support.


\end{document}